\documentstyle[11pt]{article}
\input psfig.sty

\font\tencsc=cmcsc10

\def\ra{\rightarrow}

\def\lgl{\langle}
\def\rgl{\rangle}

\def\es{\emptyset}
\def\fl{\forall}
\def\ify{\infty}
\def\op{\oplus}
\def\ot{\otimes}
\def\ov{\overline}
\def\part{\partial}
\def\sbs{\subset}
\def\sm{\simeq}
\def\ts{\times}
\def\ud{\underline}
\def\wt{\widetilde}

\def\a{\alpha}
\def\d{\delta}
\def\D{\Delta}
\def\g{\gamma}
\def\G{\Gamma}
\def\lb{\lambda}
\def\Lb{\Lambda}
\def\s{\sigma}
\def\ve{\varepsilon}
\def\vp{\varphi}

\font\tenbb=msbm10
\font\sevenbb=msbm7
\font\fivebb=msbm5
\newfam\bbfam
\textfont\bbfam=\tenbb \scriptfont\bbfam=\sevenbb
\scriptscriptfont\bbfam=\fivebb
\def\bb{\fam\bbfam}

\def\Cb{{\bb C}}
\def\Qb{{\bb Q}}
\def\Zb{{\bb Z}}

\def\Ac{{\cal A}}
\def\Hc{{\cal H}}
\def\Lc{{\cal L}}
\def\Nc{{\cal N}}
\def\Oc{{\cal O}}
\def\Sc{{\cal S}}
\def\Uc{{\cal U}}

\def\xx{\vrule height 0.5em depth 0.2em width 0.5em}

\catcode`\@=11
\def\displaylinesno #1{\displ@y\halign{
\hbox to\displaywidth{$\@lign\hfil\displaystyle##\hfil$}&
\llap{$##$}\crcr#1\crcr}}

\def\ldisplaylinesno #1{\displ@y\halign{
\hbox to\displaywidth{$\@lign\hfil\displaystyle##\hfil$}&
\kern-\displaywidth\rlap{$##$}
\tabskip\displaywidth\crcr#1\crcr}}
\catcode`\@=12

\def\sbsneq{_{\sbs \atop \neq}}

\def\semi{\mathop{>\!\!\!\triangleleft}}

\def\build#1_#2^#3{\mathrel{
\mathop{\kern 0pt#1}\limits_{#2}^{#3}}}

\begin{document}

\title{\bf Renormalization in quantum field theory and the Riemann-Hilbert problem I:
the Hopf algebra structure of graphs and the main theorem}
\author{Alain Connes and Dirk Kreimer\\[6mm]
{\small Institut des Hautes Etudes Scientifiques}\\ {\small
connes@ihes.fr, kreimer@ihes.fr}}
\date{December 1999}

\maketitle

{\small \noindent {\bf Abstract }
 {\it This paper gives a complete selfcontained proof of our result announced
 in \cite{CK} showing that renormalization in quantum field theory is
a special instance of a general mathematical procedure of extraction
 of finite values based on the Riemann-Hilbert problem.
\smallskip

\noindent We shall first show that for any quantum field theory, the
combinatorics of Feynman graphs gives rise to a Hopf algebra $\Hc$ which
is commutative as an algebra. It is the dual Hopf algebra of the
envelopping algebra of a Lie algebra $\ud G$ whose basis is labelled by
the one particle irreducible Feynman graphs. The Lie bracket of two such
graphs is computed from insertions of one graph in the other and vice
versa. The corresponding Lie group $G$ is the group of characters of
$\Hc$.

\smallskip

\noindent We shall then show that, using dimensional regularization, the
bare (unrenormalized) theory gives rise to a loop
$$
\g (z) \in G \ , \qquad z \in C
$$
where $C$ is a small circle of complex dimensions around the integer
dimension $D$ of space-time.
\noindent Our main result is that the renormalized theory is just the
evaluation at $z = D$ of the holomorphic part $\g_+$ of the Birkhoff
decomposition of $\g$.

\smallskip

\noindent We begin to analyse the group $G$ and show that it is a
semi-direct product of an easily understood abelian group by a
highly non-trivial group closely tied up with groups of
diffeomorphisms. The analysis of this latter group as well as the
interpretation of the renormalization group and of anomalous
dimensions are the content of our second paper with the same
overall title. }}

\section{Introduction}
This paper gives a complete selfcontained proof of our result announced
 in \cite{CK} showing that renormalization in quantum field theory is
a special instance of a general mathematical procedure of extraction
 of finite values based on the Riemann-Hilbert problem. In order that
 the paper be readable by non-specialists we shall begin by
giving a short introduction to both topics of renormalization and
of the Riemann-Hilbert problem, with our apologies to specialists
 in both camps for recalling well-known material.

Perturbative renormalization is by far the most successful
technique for computing physical quantities in quantum field
theory. It is well known for instance that it accurately predicts
the first ten decimal places of the anomalous magnetic moment of
the electron.

\smallskip

\noindent The physical motivation behind the renormalization technique
is quite clear and goes back to the concept of effective mass  in
nineteen century hydrodynamics. Thus for instance when applying
Newton's law,
$$
F = m \, a \leqno(1)
$$
to the motion of a spherical rigid balloon B, the inertial mass $m$ is
not the mass $m_0$ of B but is modified to
$$
m = m_0 + {\textstyle \frac{1}{2}} \, M \leqno(2)
$$
where $M$ is the mass of the volume of air occupied by B.

\smallskip

\noindent It follows for instance that the initial acceleration $a$ of
B is given, using the Archimedean law, by
$$
-Mg = \left( m_0 + {\textstyle \frac{1}{2}} \, M \right) a \leqno(3)
$$
and is always of magnitude less than $2g$.

\smallskip

\noindent The additional inertial mass $\d \, m = m - m_0$ is due to
the interaction of B with the surounding field of air and if
this interaction could not be turned off there would be no way to
measure the mass $m_0$.

\smallskip

\noindent The analogy between hydrodynamics and electromagnetism led
(through the work of Thomson, Lorentz, Kramers$\ldots$ \cite{MD}) to the
crucial distinction between the bare parameters, such as $m_0$, which
enter the field theoretic equations, and the observed parameters, such
as the inertial mass $m$.

\smallskip

\smallskip

\noindent A quantum field theory in $D=4$ dimensions, is given by
a classical action functional, $$ S \, (A) =  \int \Lc \, (A) \,
d^4 x \leqno(4) $$ where $A$ is a classical field and the
Lagrangian is of the form, $$ \Lc \, (A) =  (\part A)^2 / 2 -
\frac{m^2}{2} \, A^2 + \Lc_{\rm int} (A) \leqno(5) $$ where
$\Lc_{\rm int} (A)$ is usually a polynomial in $A$ and possibly
its derivatives.

\smallskip

\noindent One way to describe the quantum fields $\phi (x)$, is by
means of the time ordered Green's functions,
$$
G_N (x_1 , \ldots , x_N) = \lgl \, 0 \, \vert T \, \phi (x_1) \ldots
\phi (x_N) \vert \, 0 \, \rgl \leqno(6)
$$
where the time ordering symbol $T$ means that the $\phi (x_j)$'s are
written in order of increasing time from right to left.

\smallskip

\noindent The probability amplitude of a classical field configuration
$A$ is given by,
$$
e^{i \, \frac{S(A)}{\hbar}} \leqno(7)
$$
and if one could ignore the renormalization problem, the Green's
functions would be computed as,
$$
G_N (x_1 , \ldots , x_N) = \Nc \int e^{i \, \frac{S(A)}{\hbar}} \ A (x_1)
\ldots A (x_N) \, [dA] \leqno(8)
$$
where $\Nc$ is a normalization factor required to ensure the
normalization of the vacuum state,
$$
\lgl \, 0 \mid 0 \, \rgl = 1 \, . \leqno(9)
$$
It is customary to denote by the same symbol the quantum field $\phi
(x)$ appearing in (6) and the classical field $A(x)$ appearing
in the functional integral. No confusion arises from this abuse of
notation.

\smallskip

\noindent If one could ignore renormalization, the functional integral
(8) would be easy to compute in perturbation theory, i.e. by
treating the term $\Lc_{\rm int}$ in (5) as a perturbation of
$$
\Lc_0 (\phi) = (\part \phi)^2 / 2 - \frac{m^2}{2} \, \phi^2 \, .
\leqno(10)
$$
With obvious notations the action functional splits as
$$
S (\phi) = S_0 (\phi) + S_{\rm int} (\phi) \leqno(11)
$$
where the free action $S_0$ generates a Gaussian measure $\exp \, (i
\, S_0 (\phi)) \, [d \phi] = d \mu$.

\smallskip

\noindent The series expansion of the Green's functions is then given
in terms of Gaussian integrals of polynomials as,
$$
\ldisplaylinesno{
G_N (x_1 , \ldots , x_N) = \left( \sum_{n=0}^{\ify} \, i^n / n! \int
\phi (x_1) \ldots \phi (x_N) \, (S_{\rm int} (\phi))^n \, d \mu
\right) \cr
\quad \left( \sum_{n=0}^{\ify} \, i^n / n! \int S_{\rm int} (\phi)^n \, d
\mu \right)^{-1} \, . &(12) \cr
}
$$
The various terms of this expansion are computed using integration by
parts under the Gaussian measure $\mu$. This generates a large number
of terms $U(\G)$, each being labelled by a Feynman graph $\G$, and
having a numerical value $U(\G)$ obtained as a multiple integral in a
finite number of space-time variables. We shall come back later in much
detail to the precise definition of Feynman graphs and of the
corresponding integrals. But we now know enough to formulate the
problem of renormalization. As a rule the unrenormalized values
$U(\G)$ are given by nonsensical divergent integrals.

\smallskip

\noindent The conceptually really nasty divergences are called
ultraviolet\footnote{The challenge posed by the infrared problem
is formidable though. Asymptotic expansions in its presence quite
generally involve decompositions of singular expression similar to
the methods underlying renormalization theory. One can reasonably
hope that in the future singular asymptotic expansions will be
approachable by the methods advocated here.} and are associated to
the presence of arbitrarily large frequencies or equivalently to
the unboundedness of momentum space on which integration has to be
carried out. Equivalently, when one attempts to integrate in
coordinate space, one confronts divergences along diagonals,
reflecting the fact that products of field operators are defined
only on the configuration space of distinct spacetime points.

\smallskip

\noindent The physics resolution of this problem is obtained by
first introducing a cut-off in momentum space (or any suitable
regularization procedure) and then by cleverly making use of the
unobservability of the bare parameters, such as the bare mass
$m_0$. By adjusting, term by term of the perturbative expansion,
the dependence of the bare parameters on the cut-off parameter, it
is possible for a large class of theories, called renormalizable,
to eliminate the unwanted ultraviolet divergences. This resolution
of divergences can actually be carried out at the level of
integrands, with suitable derivatives with respect to external
momenta, which is the celebrated BPHZ approach to the problem.

\smallskip

\noindent The soundness of this physics resolution of the problem
makes it doubtful at first sight that it could be tied up to central
parts of mathematics.

\smallskip

\noindent It was recognized quite long ago \cite{EG} that
distribution theory together with locality were providing a
satisfactory formal approach to the problem when formulated in
configuration space, in terms of the singularities of (6) at
coinciding points, formulating the BPHZ recursion in configuration
space. The mathematical program of constructive quantum field
theory \cite{GJ} was first completed for superrenormalizable
 models, making contact with the deepest parts of hard classical analysis
 through the phase space localization and renormalization group methods.
This led to the actual rigorous mathematical construction
 of renormalizable models such as the Gross-Neveu model in 2-dimensions \cite{GK}.
 The discovery of asymptotic freedom,
 which allows to guess the asymptotic expansion of the bare parameters in terms of the
cut-off leads to the partially fulfilled hope that the rigorous contruction can be
completed for physically important theories such as QCD.

\smallskip

\noindent However neither of these important progresses does shed light on the actual
complicated combinatorics which has been successfully used by particle physicists for
 many decades to extract finite results from divergent Feynman graphs, and which is the
essence of the experimentally confirmed predictive power of quantum field theory.

\smallskip

\noindent We shall fill this gap in the present paper by unveiling
the true nature of this seemingly complicated combinatorics and by
showing that it is a special case of a general extraction of
finite values based on the Riemann-Hilbert problem.

\smallskip

\noindent Our result was announced in \cite{CK} and relies on
several previous papers \cite{CK1,K1,K2,K3,K4} but we shall give
below a complete account of its proof.
\smallskip

\noindent  The Riemann-Hilbert problem comes
from Hilbert's $21^{\rm st}$ problem which he formulated as
follows:

\begin{itemize}
\item[] ``Prove that there always exists a Fuchsian linear
differential equation with given singularities and given monodromy.''
\end{itemize}

\noindent In this form it admits a positive answer due to Plemelj
and Birkhoff (cf.~\cite{B} for a careful exposition). When
formulated in terms of linear systems of the form, $$ y'(z) = A(z)
\, y(z) \ , \ A(z) = \sum_{\a \in S} \ \frac{A \, \a }{z - \a} \,
, \leqno (13) $$ where $S$ is the given finite set of
singularities, $\ify \not\in S$, the $A_{\a}$ are complex matrices
such that $$ \sum \ A_{\a} = 0 \leqno (14) $$ to avoid
singularities at $\ify$, the answer is not always positive
\cite{Bo}, but the solution exists when the monodromy matrices
$M_{\a}$ (fig.\ref{fig1}) are sufficiently close to 1. It can then
be explicitly written as a series of polylogarithms \cite{LD}.

$$ \hbox{ \psfig{figure=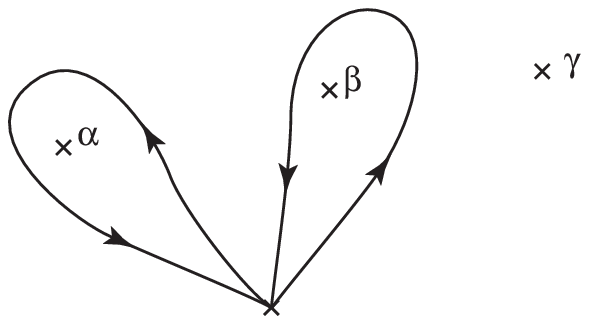} } \hbox{Figure 1} \label{fig1}
$$

\noindent Another formulation of the Riemann-Hilbert problem,
intimately tied up to the classification of holomorphic vector
bundles on the Riemann sphere $P_1 (\Cb)$, is in terms of the
Birkhoff decomposition $$ \g \, (z) = \g_- (z)^{-1} \, \g_+ (z)
\qquad z \in C \leqno (15) $$ where we let $C \sbs P_1 (\Cb)$ be a
smooth simple curve, $C_-$ the component of the complement of $C$
containing $\ify \not\in C$ and $C_+$ the other component. Both
$\g$ and $\g_{\pm}$ are loops with values in ${\rm GL}_n (\Cb)$,
$$ \g \, (z) \in G = {\rm GL}_n (\Cb) \qquad \fl \, z \in \Cb
\leqno (16) $$ and $\g_{\pm}$ are boundary values of holomorphic
maps (still denoted by the same symbol)
$$ \g_{\pm} : C_{\pm} \ra {\rm GL}_n (\Cb) \, .
\leqno (17) $$
The normalization condition $\g_- (\ify) = 1$
ensures that, if it exists, the decomposition (15) is unique
(under suitable regularity conditions).

\smallskip

\noindent The existence of the Birkhoff decomposition (15) is
equivalent to the vanishing,
$$
c_1 \, (L_j) = 0 \leqno (18)
$$
of the Chern numbers $n_j = c_1 \, (L_j)$ of the holomorphic line
bundles of the Birkhoff-Grothendieck decomposition,
$$
E = \op \, L_j \leqno (19)
$$
where $E$ is the holomorphic vector bundle on $P_1 (\Cb)$ associated to
$\g$, i.e. with total space:
$$
(C_+ \ts \Cb^n)\cup_{\g} (C_- \ts \Cb^n) \, . \leqno (20)
$$
The above discussion for $G = {\rm GL}_n (\Cb)$ extends to arbitrary
complex Lie groups.

\smallskip

\noindent When $G$ is a simply connected nilpotent complex Lie group the
existence (and uniqueness) of the Birkhoff decomposition (15)
is valid for any $\g$. When the loop $\g : C \ra G$ extends to a
holomorphic loop: $C_+ \ra G$, the Birkhoff decomposition is given by
$\g_+ = \g$, $\g_- = 1$. In general, for $z \in C_+$ the evaluation,
$$
\g \ra \g_+ (z) \in G \leqno (21)
$$
is a natural principle to extract a finite value from the singular
expression $\g (z)$. This extraction of finite values coincides with the
removal of the pole part when $G$ is the additive group $\Cb$ of complex
numbers and the loop $\g$ is meromorphic inside $C_+$ with $z$ as its
only singularity.

\smallskip

\noindent We shall first show that for any quantum field theory, the
combinatorics of Feynman graphs gives rise to a Hopf algebra $\Hc$ which
is commutative as an algebra. It is the dual Hopf algebra of the
envelopping algebra of a Lie algebra $\ud G$ whose basis is labelled by
the one particle irreducible Feynman graphs. The Lie bracket of two such
graphs is computed from insertions of one graph in the other and vice
versa. The corresponding Lie group $G$ is the group of characters of
$\Hc$.

\smallskip

\noindent We shall then show that, using dimensional regularization, the
bare (unrenormalized) theory gives rise to a loop
$$
\g (z) \in G \ , \qquad z \in C \leqno (22)
$$
where $C$ is a small circle of complex dimensions around the integer
dimension $D$ of space-time

$$ \hbox{ \psfig{figure=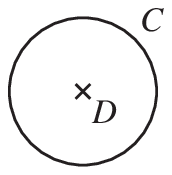} } $$

\noindent Our main result is that the renormalized theory is just the
evaluation at $z = D$ of the holomorphic part $\g_+$ of the Birkhoff
decomposition of $\g$.

\smallskip

\noindent We begin to analyse the group $G$ and show that it is a
semi-direct product of an easily understood abelian group by a
highly non-trivial group closely tied up with groups of
diffeomorphisms. The analysis of this latter group as well as the
interpretation of the renormalization group and of anomalous
dimensions are the content of our second paper with the same
overall title \cite{CKnew}.

\section{The Hopf algebra of Feynman graphs}
The Hopf algebra structure of perturbative quantum field theory is
by now well established \cite{CK1,K1,K2,K3,K4}. To the
practitioner its most exciting aspect is arguably that it is
represented as a Hopf algebra of decorated rooted trees
\cite{CK1,K3,K4} or, equivalently, as a Hopf algebra of
parenthesized words on an alphabet provided by the skeleton
expansion of the theory \cite{K1}. The relation to rooted trees
exhibits most clearly the combinatorics imposed on Feynman graphs
so that overlapping subdivergences can be resolved to deliver
local counterterms \cite{K2}. The Hopf algebra structure can be
directly formulated on graphs though \cite{K2,K4}. It is this
latter representation to which we will turn here, to make the
contact with the notation of Collins' textbook as close as
possible.

 Feynman graphs $\G$ are combinatorial
labels for the terms of the expansion (12) of Green's functions in
a given quantum field theory. They are graphs consisting of
vertices joined by lines. The vertices are of different kinds
corresponding to terms in the Lagrangian of the theory.

\smallskip

\noindent We shall require that the theory we start with is renormalizable
and include all corresponding vertices in the diagrams.

\smallskip

\noindent Thus, and if we start, for notational
simplicity,\footnote{Our results extend in a straightforward
manner to any theory renormalizable by local counterterms.}
 with $\vp^3$ in $D=6$ dimensions, we
shall have three kinds of vertices:\footnote{In the case of a
massless theory, there will be only two kinds, as the two-line
vertex corresponding to the $\vp^2$ term is missing.}

\smallskip

\noindent the three line vertex $ \hbox{ \psfig{figure=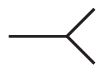}
} $ corresponding to the $\vp^3$ term of the Lagrangian;

\smallskip

\noindent the two line vertex $ \hbox{ \psfig{figure=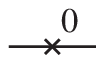} }
$ corresponding to the $\vp^2$ term of the Lagrangian;

\smallskip

\noindent the two line vertex $ \hbox{ \psfig{figure=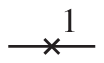} }
$ corresponding to the $(\part \, \vp)^2$ term of the Lagrangian.

\smallskip

\noindent In general the number of lines attached to a vertex is the degree
of the corresponding monomial in the Lagrangian of the theory. A line
joining two vertices is called internal. The others are attached to only
one vertex and are called external. To specify a Feynman graph one needs to
specify the values of parameters which label the external lines. When
working in configuration space these would just be the space-time points
$x_j$ of (12) associated to the corresponding external vertices
(i.e. those vertices attached to external lines). It will be more practical
to work in momentum space and to specify the external parameters of a
diagram in terms of external momenta. It is customary to orient the momenta
carried by external lines so that they all go inward. The law of
conservation of momentum means that for each connected graph $\G$,
$$
\sum \, p_i = 0 \leqno (1)
$$
where the $p_i$ are the external momenta,

$$ \hbox{ \psfig{figure=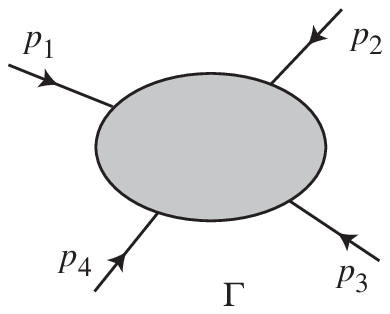} } $$

\noindent We shall use the following notation to indicate specific external
structures of a graph $\G$. We let $\G_{(0)}$ be the graph with all its
external momenta nullified, i.e.

$$ \hbox{ \psfig{figure=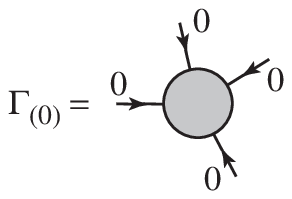} } $$

\noindent For self energy graphs, i.e. graphs $\G$ with just two
external lines, we let \footnote{in the case of a massless theory,
we take $\G_{(1)}= \left( \frac{\part}{\part \, p^2} \ \G (p)
\right)_{p^2=\mu^2} $ } $$ \G_{(1)} = \left( \frac{\part}{\part \,
p^2} \ \G (p) \right)_{p=0} \leqno (2) $$ where $p$ is the
momentum flowing through the diagram. Note that the sign of $p$ is
irrelevant in (2). This notation might seem confusing at first
sight but becomes clear if one thinks of the external structure of
graphs in terms of distributions. This is necessary to have
space-time parameters $x_j$ on the same footing as the momentum
parameters $p_j$ using Fourier transform. We shall return to this point
 in greater detail after the proof of theorem 1.

\smallskip

\noindent A Feynman graph $\G$ is called one particle irreducible
(1PI) if it is connected and cannot be disconnected by removing a
single line. The following graph is one particle reducible:

$$ \hbox{ \psfig{figure=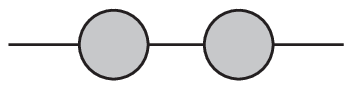} } $$

\noindent The diagram $ \hbox{ \psfig{figure=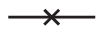} } $ is
considered as not 1PI.\footnote{This is a conveniently chosen  but
immaterial convention.}

\smallskip

\noindent Let us now define the Hopf algebra $\Hc$.\footnote{We work with
any field of coefficients such as $\Qb$ or $\Cb$.} As a linear space $\Hc$
has a basis labelled by all Feynman graphs $\G$ which are disjoint unions
of 1PI graphs.
$$
\G = \bigcup_{j=1}^{n} \G_j \, . \leqno (3)
$$
The case $\G = \es$ is allowed.

\smallskip

\noindent The product in $\Hc$ is bilinear and given on the basis by the
operation of disjoint union:
$$
\G \cdot \G' = \G \cup \G' \, . \leqno (4)
$$
It is obviously commutative and associative and it is convenient to use the
multiplicative notation instead of $\cup$. In particular one lets 1 denote the
graph $\G = \es$, it is the unit of the algebra $\Hc$. To define the
coproduct,
$$
\D : \Hc \ra \Hc \ot \Hc \leqno (5)
$$
it is enough, since it is a homomorphism of algebras, to give it on the
generators of $\Hc$, i.e. the 1PI graphs. We let,
$$
\D \G = \G \ot 1 + 1 \ot \G + \sum_{\g \sbsneq \G} \g_{(i)} \ot \G /
\g_{(i)} \leqno (6)
$$
where the notations are as follows.

\smallskip

\noindent First $\g$ is a non trivial\footnote{i.e. non empty and with non
empty complement.} subset $\g \sbs \G^{(1)}$ of the set of internal lines
of $\G$ whose connected components $\g'$ are 1PI and fulfill the following
condition:
$$
\ldisplaylinesno{
\hbox{The set $\ve (\g')$ of lines\footnote{} of $\G$ which meet $\g'$}
&(7) \cr
\hbox{without belonging to $\g'$, has two or three elements.} \cr
}
$$\footnotetext{Internal or external lines.}
We let $\g'_{(i)}$ be the Feynman graph with $\g'$ as set of internal
lines, $\ve \, (\g')$ as external lines, and external structure given by
nullified external momenta for $i=0$ and by (2) if $i=1$.

\smallskip

\noindent In the sum (6) the multi index $i$ has one value for
each connected component of $\g$. This value is 0 or 1 for a self
energy component and 0 for a vertex. One lets $\g_{(i)}$ be the
product of the graphs $\g'_{(i)}$ corresponding to the connected
components of $\g$. The graph $\G / \g_{(i)}$ is obtained by
replacing each of the connected components $\g'$ of $\g$ by the
corresponding vertex, with the same label $(i)$ as
$\g'_{(i)}$.\footnote{One checks that $\G / \g (i)$ is still a 1PI
graph.} The sum (6) is over all values of the multi index $i$. It
is important to note that though the components $\g'$ of $\g$ are
pairwise disjoint by construction, the 1PI graphs $\g'_{(i)}$ are
not necessarily disjoint since they can have common external legs
$\ve \, (\g') \cap \ve \, (\g'') \ne \es$.

\smallskip

\noindent This happens for instance in the following example,

$$ \hbox{ \psfig{figure=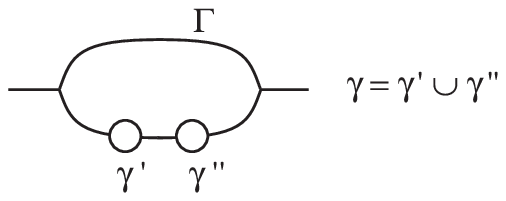} } \leqno (8) $$

\noindent Note also that the external structure of $\G / \g_{(i)}$ is
identical to that of $\G$.

\smallskip

\noindent To get familiar with (6) we shall now give a few
examples of coproducts.

$$ \hbox{ \psfig{figure=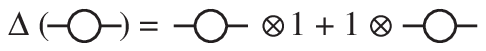} } \leqno (9) $$

$$ \hbox{ \psfig{figure=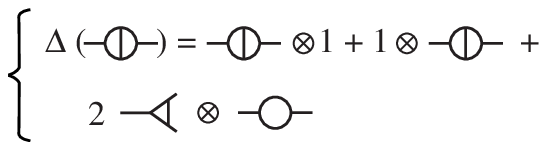} } \leqno (10) $$

$$ \hbox{ \psfig{figure=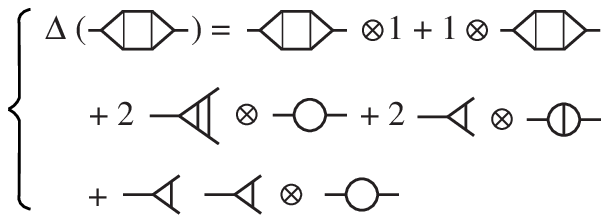} } \leqno (11) $$

$$ \hbox{ \psfig{figure=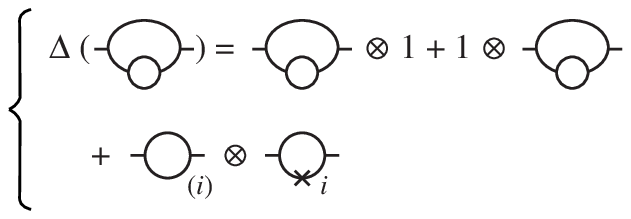} } \leqno (12) $$

\noindent where one sums over $i=0,1$.

\smallskip

\noindent Example (9) shows what happens in general for graphs
with no subgraph fulfilling (7), i.e. graphs without
subdivergences. Example (10) shows an example of overlapping
subdivergences. Example (11) also, but it illustrates an important
general feature of the coproduct $\D (\G)$ for $\G$ a 1PI graph,
namely that $\D (\G) \in \Hc \ot \Hc_{(1)}$ where $\Hc_{(1)}$ is
the subspace of $\Hc$ generated by 1 and the 1PI graphs. Similar
examples were given in \cite{CK1,K2}.

\smallskip

\noindent The coproduct $\D$ defined by (6) on 1PI graphs extends
uniquely to a homomorphism from $\Hc$ to $\Hc \ot \Hc$. The main
result of this section is (\cite{K1,K2}):

\medskip

\noindent {\tencsc Theorem 1.} {\it The pair $(\Hc , \D)$ is a
Hopf algebra.}

\medskip

\noindent {\it Proof.} Our first task will be to prove that $\D$ is
coassociative, i.e. that,
$$
(\D \ot {\rm id}) \, \D = ({\rm id} \ot \D) \, \D \, . \leqno (13)
$$
Since both sides of (13) are algebra homomorphisms from $\Hc$ to
$\Hc \ot \Hc \ot \Hc$, it will be enough to check that they give the same
result on 1PI graphs $\G$. Thus we fix the 1PI graph $\G$ and let
$\Hc_{\G}$ be the linear subspace of $\Hc$ spanned by 1PI graphs with the
same external structure as $\G$. Also we let $\Hc_c$ be the subalgebra of
$\Hc$ generated by the 1PI graphs $\g$ with two or three external legs and
with external structure of type $(i)$, $i = 0,1$. One has by (6)
that
$$
\D \, \G - \G \ot 1 \in \Hc_c \ot \Hc_{\G} \leqno (14)
$$
while
$$
\D \, \Hc_c \sbs \Hc_c \ot \Hc_c \, . \leqno (15)
$$
Thus we get,
$$
(\D \ot {\rm id}) \, \D \, \G - \D \, \G \ot 1 \in \Hc_c \ot \Hc_c \ot
\Hc_{\G} \, . \leqno (16)
$$
To be more specific we have the formula,
$$
(\D \ot {\rm id}) \, \D \, \G - \D \, \G \ot 1 = \sum_{\g \sbsneq \G} \D \,
\g_{(i)} \ot \G / \g (i) \leqno (17)
$$
where $\g = \es$ is allowed now in the summation of the right hand side.

\smallskip

\noindent We need a nice formula for $\D \, \g_{(i)}$ which is defined as
$$
\Pi \, \D \, \g_{(i_j)}^j \leqno (18)
$$
where the $\g^j$ are the components of $\g$.

\smallskip

\noindent For each of the graphs $\g'' = \g_{(i_j)}^j$ the formula
(6) for the coproduct simplifies to give
$$
\D \, \g'' = \sum_{\g' \sbs \g''} \g'_{(k)} \ot \g'' / \g'_{(k)} \leqno
(19)
$$
where the subset $\g'$ of the set $\g''^{(1)}$ of internal lines of $\g''$
is now allowed to be empty (which gives $1 \ot \g''$) or full (which gives
$\g'' \ot 1$). Of course the sum (19) is restricted to $\g'$ such
that each component is 1PI and satisfies (7) relative to $\g'' =
\g_{(i_j)}^j$. But this is equivalent to fulfilling (7) relative
to $\G$ since one has,
$$
\ve_{\g''} (\g') = \ve_{\G} (\g') \, . \leqno (20)
$$
Indeed since $\g''$ is a subgraph of $\G$ one has $\ve_{\g''} (\g') \sbs
\ve_{\G} (\g')$ but every line $\ell \in \ve_{\G} (\g')$ is a line of $\G$
which meets $\g''^{(1)}$ and hence belongs to $\g''$ which gives equality.

\smallskip

\noindent The equality (20) also shows that the symbol
$\g'_{(k)}$ in (19) can be taken relative to the full graph $\G$.
We can now combine (17) (18) and (19) to write
the following formula:
$$
(\D \ot {\rm id}) \, \D \, \G - \D \, \G \ot 1 = \sum_{\g' \sbs \g \sbsneq
\G} \g'_{(k)} \ot \g_{(i)} / \g'_{(k)} \ot \G / \g_{(i)} \, , \leqno (21)
$$
where both $\g$ and $\g'$ are subsets of the set of internal lines
$\G^{(1)}$ of $\G$ and $\g'$ is a subset of $\g$. Both $\g$ and $\g'$
satisfy (7). Since $\g$ is not necessarily connected we need to
define the symbol $\g_{(i)} / \g'_{(k)}$ as the replacement in the not
necessarily connected graph $\g_{(i)}$ of each of the component of $\g'$ by
the corresponding vertex. The only point to remember is that if a component
of $\g'$ is equal to a component of $\g$ the corresponding index $k_j$ is
equal to $i_j$ (following (19)) and the corresponding term in $\g
/ \g'$ is equal to 1.

$$ \hbox{ \psfig{figure=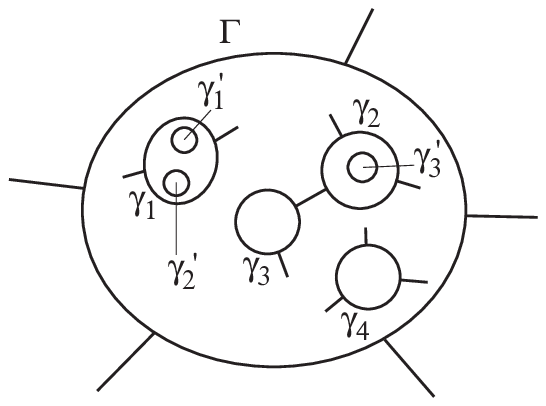} } $$

\noindent Let us now compute $({\rm id} \ot \D) \, \D \, \G - \D \, \G \ot
1$ starting from the equality,
$$
\D \, \G = \G \ot 1 + \sum_{\g' \sbsneq \G} \g'_{(k)} \ot \G / \g'_{(k)}
\leqno (22)
$$
where, unlike in (6), we allow $\g' = \es$ in the sum. Let us
define $\D' : \Hc \ra \Hc \ot \Hc$ by
$$
\D' \, X =  \D \, X - X \ot 1 \qquad \fl \, X \in \Hc \, . \leqno (23)
$$
One has $({\rm id} \ot \D') \, \D' \, X = ({\rm id} \ot \D') \, \D \, X$
since $\D' \, 1 = 0$, and $({\rm id} \ot \D') \, \D \, X = ({\rm id} \ot \D
) \, \D \, X - ({\rm id} \ot {\rm id} \ot 1) \, \D \, X = ({\rm id} \ot \D )
\, \D \, X - \D \, X \ot 1$. Thus,
$$
({\rm id} \ot \D) \, \D \, \G - \D \, \G \ot 1 = ({\rm id} \ot \D') \, \D'
\, \G \, . \leqno (24)
$$
Moreover (22) gives the formula for $\D'$ on $\Hc_{(1)}$ which is
thus enough to get,
$$
({\rm id} \ot \D) \, \D \, \G - \D \, \G \ot 1 = \sum_{\g' , \g''} \g'_{(k)}
\ot \g''_{(j)} \ot (\G / \g'_{(k)}) / \g''_{(j)} \, . \leqno (25)
$$
In this sum $\g'$ varies through the (possibly empty) admissible subsets of
$\G^{(1)}$, $\g' \ne \G^{(1)}$, while $\g''$ varies through the (possibly
empty) admissible subsets of the set of lines $\G'^{(1)}$ of the graph
$$
\G' = \G / \g'_{(k)} \, . \leqno (26)
$$
To prove that the sums (21) and (25) are equal it is
enough to prove that for any admissible subset $\g' \, {_{\textstyle \sbs}
\atop \ne} \, \G^{(1)}$ the
corresponding sums are equal. We also fix the multi index $k$. The sum
(25) then only depends upon the graph $\G' = \G / \g'_{(k)}$. We
thus need to show the equality
$$
\sum_{ \g' \sbs \g \sbsneq \G} \g_{(i)} / \g'_{(k)} \ot \G / \g_{(i)} =
\sum_{\g''} \g''_{(j)} \ot \G' / \g''_{(j)} \, . \leqno (27)
$$
Let $\pi : \G \ra \G'$ be the continuous projection,
$$
\pi : \G \ra \G / \g'_{(k)} = \G' \, . \leqno (28)
$$
Let us show that the map $\rho$ which associates to every admissible subset
$\g$ of $\G$ containing $\g'$ its image $\g / \g'_{(k)}$ in $\G'$ gives a
bijection with the admissible subsets $\g''$ of $\G'$.

\smallskip

\noindent Since each connected component of $\g'$ is contained in a
connected component of $\g$, we see that $\rho \, (\g)$ is admissible in
$\G'$. Note that when a connected component $\g^0$ of $\g$ is equal to a
connected component of $\g'$ its image in $\G'$ is defined to be empty. For
the components of $\g$ which are not components of $\g'$ their image is just
the contraction $\g / \g'_{(k)}$ by those components of $\g'$ which it
contains. This does not alter the external leg structure, so that
$$
\g_{(i)}^0 / \g'_{(k)} = (\g^0 / \g'_{(k)})_{(i)} \, . \leqno (29)
$$
The inverse of the map $\rho$ is obtained as follows. Given $\g''$ an
admissible subset of $\G'$, one associates to each component of $\g''$ its
inverse image by $\pi$ which gives a component of $\g$. Moreover to each
vertex $v$ of $\G'$ which does not belong to $\g''$ but which came from the
contraction of a component of $\g'$, one associates this component as a new
component of $\g$. It is clear then that $\g' \sbs \g \, {_{\textstyle \sbs}
\atop \ne} \, \G$ and that
$\g$ is admissible in $\G$ with $\rho \, (\g) = \g''$.

\smallskip

\noindent To prove (27) we can thus fix $\g$ and $\g'' = \rho \,
(\g)$. For those components of $\g$ equal to components of $\g'$ the index
$i_0$ is necessarily equal to $k_0$ so they contribute in the same way to
both sides of (27). For the other components there is freedom in
the choice of $i_0$ or $j_0$ but using (29) the two contributions
to (27) are also equal. This ends the proof of the coassociativity
of the coproduct $\D$ and it remains to show that the bialgebra $\Hc$ admits
an antipode.

\smallskip

\noindent This can be easily proved by induction but it is worthwile to
discuss various gradings of the Hopf algebra $\Hc$ associated to natural
combinatorial features of the graphs. Any such grading will associate an
integer $n \, (\G) \in \Zb$ to each 1PI graph, the corresponding grading of
the algebra $\Hc$ is then given by
$$
\deg \, (\G_1 \ldots \G_e) = \sum n \, (\G_j) \ , \quad \deg \, (1) = 0
\leqno (30)
$$
and the only interesting property is the compatibility with the coproduct
which means, using (6), that
$$
\deg \, (\g) + \deg \, (\G / \g) = \deg \, (\G) \leqno (31)
$$
for any admissible $\g$.

\smallskip

\noindent The first two natural gradings are given by
$$
I \, (\G) = \hbox{number of internal lines in} \ \G \leqno (32)
$$
and
$$
v \, (\G) = V \, (\G) - 1 = \hbox{number of vertices of} \ \G - 1 \, .
\leqno (33)
$$
An important combination of these two gradings is
$$
L = I - v = I - V + 1 \leqno (34)
$$
which is the loop number of the graph or equivalently the rank of its first
homology group. It governs the power of $\hbar$ which appears in the
evaluation of the graph.

\smallskip

\noindent Note that the number of external lines of a graph is not a good
grading since it fails to fulfill (31). For any of the three
gradings $I$, $v$, $L$ one has the following further compatibility with the
Hopf algebra structure of $\Hc$,

\medskip

\noindent {\tencsc Lemma.} a) {\it The scalars are the only
homogeneous elements of degree $0$.}

\smallskip

\noindent b) {\it For any non scalar homogeneous $X \in \Hc$ one has
$$
\D \, X = X \ot 1 + 1 \ot X + \sum X' \ot X''
$$
where $X'$, $X''$ are homogeneous of degree strictly less than the degree of
$X$.}

\medskip

\noindent {\it Proof.} a) Since the diagram $ \hbox{
\psfig{figure=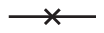} } $ is excluded we see that $I \, (\G) =
0$ or $v \, (\G) = 0$ is excluded. Also if $L \, (\G) = 0$ the
$\G$ is a tree but a tree diagram cannot be 1PI unless it is equal
to $ \hbox{ \psfig{figure=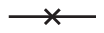} } $ which is excluded.

\smallskip

\noindent b) Since $X$ is a linear combination of homogeneous monomials it
is enough to prove b) for $\Pi \, \G_i$ and in fact for 1PI graphs $\G$.
Using (6) it is enough to check that the degree of a non empty
$\g$ is strictly positive, which follows from a).~\xx

\medskip

\noindent We can now end the proof of Theorem~1 and give an inductive
formula for the antipode $S$.

\smallskip

\noindent The counit $\ov e$ is given as a character of $\Hc$ by
$$ \ov e \, (1) = 1 \, , \ \ov e \, (\G) = 0 \qquad \fl \, \G \ne
\es \, . \leqno (35) $$ The defining equation for the antipode is,
$$ m \, (S \ot {\rm id}) \, \D \, (a) = \ov e \, (a) \, 1 \qquad
\fl \, a \in \Hc \leqno (36) $$ and its existence is obtained by
induction using the formula, $$ S \, (X) = - X - \sum S \, (X') \,
X'' \leqno (37) $$ for any non scalar homogeneous $X \in \Hc$
using the notations of Lemma~2. This antipode also fulfills the
other required identity \cite{CK1,K1}, $$ m \, ( {\rm id}\ot S) \,
\D \, (a) = \ve \, (a) \, 1 \qquad \fl \, a \in \Hc. \leqno  $$

\smallskip

\noindent This completes the proof of Theorem~1.~\xx

\medskip

\noindent Let us now be more specific about the external structure
of diagrams. Given a 1PI graph, with $n$ external legs labelled by
$i \in \{ 1 , \ldots , n \}$ we specify its external structure by
giving a distribution $\s$ defined on a suitable test space $\Sc$
of smooth functions on $$ \left\{ (p_i)_{i=1 , \ldots , n} \ ; \
\sum \, p_i = 0 \right\} = E_n \, . \leqno (38) $$ Thus $\sigma$
is a continuous linear map, $$ \sigma : \Sc \, (E) \ra \Cb \, .
\leqno (39) $$ To a graph $\G$ with external structure $\sigma$ we
have associated above an element of the Hopf algebra $\Hc$ and we
require the linearity of this map, i.e. $$ \d_{(\G , \lb_1
\sigma_1 + \lb_2 \sigma_2)} = \lb_1 \, \d_{(\G , \sigma_1)} +
\lb_2 \, \d_{(\G , \sigma_2)} \, . \leqno (40) $$ One can easily
check that this relation is compatible with the coproduct.

\smallskip

\noindent There is considerable freedom in the choice of the
external structure of the 1PI graphs which occur in the left hand
side of the last term of the coproduct formula (6). We wrote the
proof of Theorem~1 in such a way that this freedom is apparent.
The only thing which matters is that, say for self energy graphs,
the distributions $\sigma_0$ and $\sigma_1$ satisfy, $$ \sigma_0 \, (a \, m^2+ b
\, p^2) = a \ , \ \sigma_1 \, (a \, m^2+ b \, p^2) = b \, , \leqno (41) $$
where $p=p_1$ is the natural coordinate on $E_2$ and $m$ is a mass parameter.

This freedom in the definition of the Hopf algebra $\Hc$ is the
same as the freedom in the choice of parametrization of the
corresponding QFT and it is important to make full use of it, for
instance for massless theories in which the above choice of
nullified external momenta is not appropriate and one would rely
only on $\sigma_1(bp^2)=b$.

\smallskip

\noindent For simplicity of exposition we shall keep the above choice, the
generalization being obvious.

\smallskip

\noindent We shall now apply the Milnor-Moore theorem to the bigraded Hopf
algebra $\Hc$. The two natural gradings are $v$ and $L$, the grading $I$
is
just their sum.

\smallskip

\noindent This theorem first gives a Lie algebra structure on the
linear space, $$ \bigoplus_{\G} \ \Sc \, (E_{\G}) = L \leqno (42)
$$ where for each 1PI graph $\G$, we let $\Sc \, (E_{\G})$ be the
test space associated to the external lines of $\G$ as in (38).
Given $X \in L$ we consider the linear form $Z_X$ on $\Hc$ given,
on the monomials $\G$, by $$ \lgl \G , Z_X \rgl = 0 \leqno (43) $$
unless $\G$ is a (connected) 1PI, in which case, $$ \lgl \G , Z_X
\rgl =\lgl \sigma_{\G} , X_{\G} \rgl \leqno (44) $$ where
$\sigma_{\G}$ is the distribution giving the external structure of
$\G$ and where $X_{\G}$ is the corresponding component of $X$. By
construction $Z_X$ is an infinitesimal character of $\Hc$ and the
same property holds for the commutator, $$ [Z_{X_1} , Z_{X_2}] =
Z_{X_1} \, Z_{X_2} - Z_{X_2} \, Z_{X_1} \leqno (45) $$ where the
product in the right hand side is given by the coproduct of $\Hc$,
i.e. by $$ \lgl Z_1 \, Z_2 , \G \rgl = \lgl Z_1 \ot Z_2 , \D \, \G
\rgl \, . \leqno (46) $$ The computation of the Lie bracket is
straightforward as in \cite{CM} p.~207 or \cite{CK1} and is given
as follows. One lets $\G_j$, $j = 1,2$ be 1PI graphs and $\vp_j
\in \Sc \, (E_{\G_j})$ corresponding test functions.

\smallskip

\noindent For $i \in \{ 0,1 \}$, we let $n_i \, (\G_1 , \G_2 ; \G)$ be the
number of subgraphs of $\G$ which are isomorphic to $\G_1$ while
$$
\G / \G_1 (i) \sm \G_2 \, . \leqno (47)
$$
We let $(\G , \vp)$ be the element of $L$ associated to $\vp \in \Sc \,
(E_{\G})$, the Lie bracket of $(\G_1 , \vp_1)$ and $(\G_2 , \vp_2)$ is then,
$$
\sum_{\G , i} \, \sigma_i \, (\vp_1) \ n_i \, (\G_1 , \G_2 ; \G) \, (\G ,
\vp_2)
- \sigma_i \, (\vp_2) \ n_i \, (\G_2 , \G_1 ; \G) \, (\G , \vp_1) \, . \leqno
(48)
$$
What is obvious in this formula is that it vanishes if $\sigma_i \, (\vp_j) =
0$ and hence we let $L_0$ be the subspace of $L$ given by,
$$
L_0 = \op \ \Sc \, (E_{\G})_0 \, , \, \Sc \, (E_{\G})_0 = \{ \vp ; \, \sigma_i
\, (\vp) = 0 \ , \ i = 0,1 \} \, . \leqno (49)
$$
It is by construction a subspace of finite codimension in each of the $\Sc
\, (E_{\G})$.

\smallskip

\noindent We need a natural supplement and in view of (41) we
should choose the obvious test functions, $$ \vp_0 \, (p) = m^2,
\, \vp_1 \, (p) = p^2 \, . \leqno (50) $$ We shall thus, for any
1PI self energy graph, let $$ (\G^{(i)}) = (\G , \vp_i) \, .
\leqno (51) $$ Similarly for a vertex graph with the constant
function 1. One checks using (48) that the $\G^{(i)}$ for 1PI
graphs with two or three external legs, do generate a Lie
subalgebra $$ L_c = \left\{ \sum \lb \, (\G^{(i)}) \right\} \, .
\leqno (52) $$ We can now state the following simple fact,

\medskip

\noindent {\tencsc Theorem 2.} {\it The Lie algebra $L$ is the semi-direct
product of $L_0$ by $L_c$. The Lie algebra $L_0$ is abelian, while $L_c$ has
a canonical basis labelled by the $\G^{(i)}$ and such that
$$
[\G , \G'] = \sum_v \, \G \circ_v \G' - \sum_{v'} \, \G' \circ_{v'} \G
$$
where $\G \circ_v \G'$ is obtained by grafting $\G'$ on $\G$ at $v$.}

\medskip

\noindent {\it Proof.} Using (48) it is clear that $L_0$ is an abelian Lie
subalgebra of $L$. The Lie bracket of $(\G^{(i)}) \in L_c$ with $(\G , \vp)
\in L_0$ is given by
$$
\sum n_i \, (\G^{(i)} , \G ; \G') \, (\G' , \vp) \leqno (53)
$$
which belongs to $L_0$ so that $[L_c , L_0] \sbs L_0$.

\smallskip

\noindent To simplify the Lie bracket of the $(\G^{(i)})$ in $L_c$
we introduce the new basis, $$ \G^{(i)} = -S \, (\G) \, (\G)
\leqno (54) $$ where $S \, (\G)$ is the symmetry factor of a
Feynman graph, i.e. the cardinality of its group of automorphisms.
In other words if another graph $\G'$ is isomorphic to $\G$ there
are exactly $S \, (\G)$ such isomorphisms. From the definition
(47) of $n_i \, (\G_1 , \G_2 ; \G)$ we see that $S \, (\G_1) \, S
\, (\G_2) \ n_i \, (\G_1 , \G_2 ; \G)$ is the number of pairs $j_1
, j_2$ of an embedding $$ j_1 : \G_1 \ra \G \leqno (55) $$ and of
an isomorphism $$ j_2 : \G_2 \sm \G / \G_1 \, (i) \, . \leqno (56)
$$ Giving such a pair is the same as giving a vertex $v$ of type
(i) of $\G_2$ and an isomorphism, $$ j : \G_2 \circ_v \G_1 \ra \G
\, . \leqno (57) $$ Since there are $S \, (\G)$ such isomorphisms
when $\G \sim \G_2 \circ_v \G_1$ we get the formula of Theorem~2
using (48).~\xx

\medskip

\noindent It is clear from Theorem~2 that the Lie algebra $L_c$ is
independent of the choice of the distributions $\sigma_j$ fulfilling (41). The
same remark applies to $L$ using (53).

\smallskip

\noindent By the Milnor-Moore theorem the Hopf algebra $\Hc$ is the dual of
the envelopping algebra $\Uc \, (L)$. The linear subspace $\Hc_{(1)}$ of
$\Hc$ spanned by 1 and the 1PI graphs give a natural system of affine
coordinates on the group $G$ of characters of $\Hc$, i.e. of homomorphisms,
$$
\vp : \Hc \ra \Cb \ , \ \vp \, (XY) = \vp \, (X) \ \vp \, (Y) \qquad \fl \,
X,Y \in \Hc \leqno (58)
$$
from the algebra $\Hc$ to complex numbers.

\smallskip

\noindent We shall only consider homomorphisms which are continuous with
respect to the distributions labelling the external stucture of graphs.

\smallskip

\noindent The group operation in $G$ is given by the convolution,
$$
(\vp_1 * \vp_2) \, (X) = \lgl \vp_1 \ot \vp_2 , \D \, X \rgl \qquad \fl \, X
\in \Hc \, . \leqno (59)
$$
The Hopf subalgebra $\Hc_c$ of $\Hc$ generated by the 1PI with two or three
external legs and external structure given by the $\sigma_i$, is the dual of
the envelopping algebra $\Uc \, (L_c)$ and we let $G_c$ be the group of
characters of $\Hc_c$.

\smallskip

\noindent The map $\vp \ra \vp / \Hc_c$ defines a group homomorphism,
$$
\rho : G \ra G_c \leqno (60)
$$
and as in Theorem~2 one has,

\medskip

\noindent {\tencsc Proposition 3.} {\it The kernel $G_0$ of $\rho$ is
abelian and $G$ is the semi-direct product $G = G_0 \semi G_c$ of $G_0$ by
the action of $G_c$.}

\medskip

\noindent {\it Proof.} A character $\vp$ of $\Hc$ belongs to $G_0$ iff its
restriction to $\Hc_c$ is the augmentation map. This just means that for any
1PI graph $\g$ one has $\vp \, (\g_{(i)}) = 0$. Thus the convolution of two
characters $\vp_j \in G_0$ is just given by,
$$
(\vp_1 * \vp_2) \, (\G) = \vp_1 \, (\G) + \vp_2 \, (\G) \leqno (61)
$$
for any 1PI graph $\G$. This determines the character $\vp_1 * \vp_2$
uniquely, so that $\vp_1 * \vp_2 = \vp_2 * \vp_1$.

\smallskip

\noindent Let us now construct a section $\rho' : G_c \ra G$ which is a
homomorphism. We just need to construct a homomorphism,
$$
\vp \ra \wt \vp = \vp \circ \rho' \leqno (62)
$$
from characters of $\Hc_c$ to characters of $\Hc$, such that
$$
\wt \vp / \Hc_c = \vp \, . \leqno (63)
$$
It is enough to extend $\vp$ to all 1PI graphs $(\G , \sigma)$ where $\sigma$ is
the external structure, this then determines $\wt \vp$ uniquely as a
character and one just has to check that,
$$
(\vp_1 * \vp_2)^{\wt{ \ }} = \wt{\vp}_1 * \wt{\vp}_2 \, . \leqno (64)
$$
We let $\wt \vp \, (\G , \sigma) = 0$ for any 1PI graph with $n \ne 2,3$
external legs or any 1PI graph with 2 or 3 external legs such that,
$$
\sigma \, ((p^2)^i) = 0 \qquad i = 0,1 \quad \hbox{if} \quad n=2 \ , \qquad i =
0 \quad \hbox{if} \quad n=3 \, . \leqno (65)
$$
By (41) this gives a natural supplement of $\Hc_{c(1)}$ in $\Hc_{(1)}$
and determines $\wt \vp$ uniquely.

\smallskip

\noindent To check (64) it is enough to test both sides on 1PI graphs $(\G ,
\sigma)$. One uses (6) to get,
$$
(\wt{\vp}_1 * \wt{\vp}_2) \, (\G , \sigma) = \wt{\vp}_1 \, (\G , \sigma) +
\wt{\vp}_2 \, (\G , \sigma) + \sum \, \vp_1 \, (\g_{(i)}) \ \wt{\vp}_2 \, (\G /
\g_{(i)} , \sigma) \, . \leqno (66)
$$
For any $(\G , \sigma)$ in the above supplement of $\Hc_{c(1)}$ the right hand
side clearly vanishes. The same holds for $(\vp_1 * \vp_2)^{\wt{ \ }} \, (\G
, \sigma)$ by construction. For any $(\G , \sigma_i) \in \Hc_{c(1)}$ one simply
gets the formula for $(\vp_1 * \vp_2) \, (\G , \sigma_i) = (\vp_1 *
\vp_2)^{\wt{ \ }} \, (\G , \sigma_i)$. This gives (64). We have thus proved
that $\rho$ is a surjective homomorphism from $G$ to $G_c$ and that $\rho' :
G_c \ra G$ is a group homomorphism such that
$$
\rho \circ \rho' = {\rm id}_{G_c} \, . \leqno (67)
$$
\hfill \xx

\medskip

Let us now compute explicitly the action of $G_c$ on $G_0$ given by inner
automorphisms,
$$
\a \, (\vp_1) \, \vp = \wt{\vp}_1 * \vp * \wt{\vp}_1^{-1} \, . \leqno (68)
$$
One has
$$
(\vp * \wt{\vp}_1^{-1}) \, (\G , \s) = \vp \, (\G , \s) \leqno (69)
$$
for any $(\G , \s)$ in the supplement of $\Hc_{c(1)}$ in $\Hc_{(1)}$. Thus,
$$
\a \, (\vp_1) \, (\vp) \, (\G , \s) = \vp \, (\G , \s) + \sum \, \vp_1 \,
(\g_{(i)}) \ \vp \, (\G / \g_{(i)} , \s) \, . \leqno (70)
$$
This formula shows that the action of $G_c$ on $G_0$ is just a linear
representation of $G_c$ on the vector group $G_0$.

\medskip

\noindent A few remarks are in order. First of all, the Hopf
algebra of Feynman graphs as presented in \cite{K2} is the
same as ${\cal H}_c$. As we have seen in theorem 2 and proposition 3,
the only nontrivial part in the Hopf algebra ${\cal H}$ is ${\cal H}_c$.
 For instance, the Birkhoff
decomposition of a loop $\g (z)
\in G$, is readily obtained from the Birkhoff
decomposition of its homomorphic image, $\g_c (z)
\in G_c$ since the latter allows to move back to a
loop $\g_0 (z)
\in G_0$ with $G_0$ abelian.
Dealing with the Hopf algebra ${\cal H}$  allows however
to treat oversubtractions and operator product
expansions in an effective manner.
\medskip

\noindent Second, one should point out that there is a deep relation between
the Hopf algebra ${\cal H}_c$ and the Hopf algebra of
rooted trees \cite{CK1,K1,K2}.
 This is essential for the
practitioner of QFT \cite{K4}. The relation was first
established using the particular structures imposed on the
perturbation series by the Schwinger-Dyson equation \cite{K1}.
Alternative reformulations of this Hopf algebra  confirmed this
relation \cite{KW} in full agreement with the general analysis
of \cite{K2}.

\noindent A full description of the relation between the Hopf
algebras of graphs and of rooted trees in the language established
here will be given in the second part of this paper \cite{CKnew}.

\section{Renormalization and the Birkhoff decomposition}

We shall show that given a renormalizable quantum field theory in $D$
space-time dimensions, the bare theory gives rise, using dimensional
regularization, to a loop $\g$ of elements in the group $G$ associated to
the theory in section 2. The parameter $z$ of the loop $\g \, (z)$ is a
complex variable and $\g \, (z)$ makes sense for $z \ne D$ in a neighborhood
of $D$, and in particular on a small circle $C$ centered at $z = D$. Our
main result is that the renormalized theory is just the evaluation at $z=D$
of the holomorphic piece $\g_+$ in the Birkhoff decomposition,
$$
\g \, (z) = \g_- \, (z)^{-1} \ \g_+ \, (z) \leqno (1)
$$
of the loop $\g$ as a product of two holomorphic maps $\g_{\pm}$ from the
respective components $C_{\pm}$ of the complement of the circle $C$ in the
Riemann sphere $\Cb \, P^1$. As in section 2 we shall, for simplicity, deal
with $\vp^3$ theory in $D=6$ dimensions since it exhibits all the important
general difficulties of renormalizable theories which are relevant here.

\smallskip

\noindent The loop $\g \, (z)$ is obtained by applying dimensional
regularization (Dim. Reg.) in the evaluation of the bare values of Feynman
graphs $\G$, and our first task is to recall the Feynman rules which
associate an integral
$$
U_{\G} \, (p_1 , \ldots , p_N) = \int d^d \, k_1 \ldots d^d \, k_L \ I_{\G}
\, (p_1 , \ldots , p_N , k_1 , \ldots , k_L) \leqno (2)
$$
to every graph $\G$.

\smallskip

\noindent For convenience we shall formulate the Feynman rules in Euclidean
space-time to eliminate irrelevant singularities on the mass shell and
powers of $i = \sqrt{-1}$. In order to write these rules directly in $d$
space-time dimensions it is important (\cite{H}) to introduce a unit of mass
$\mu$ and to replace the coupling constant $g$ which appears in the
Lagrangian as the coefficient of $\vp^3 / 3!$ by $\mu^{3-d/2} \, g$. The
effect then is that $g$ is dimensionless for any value of $d$ since the
dimension of the field $\vp$ is $\frac{d}{2} - 1$ in a $d$-dimensional
space-time.

\smallskip

\noindent The integrand $I_{\G} \, (p_1 , \ldots , p_N , k_1 ,
\ldots , k_L)$ contains $L$ internal momenta $k_j$, where $L$ is
the loop number of the graph $\G$, and results from the following
rules, $$ \hbox{Assign a factor $\frac{1}{k^2+m^2}$ to each
internal line.} \leqno (3) $$ $$ \hbox{Assign a momentum
conservation rule to each vertex.} \leqno (4) $$ $$ \hbox{Assign a
factor $\mu^{3-d/2} \, g$ to each 3-point vertex.} \leqno (5) $$
$$ \hbox{Assign a factor $m^2$ to each 2-point vertex$_{(0)}$.}
\leqno (6) $$ $$ \hbox{Assign a factor $p^2$ to each 2-point
vertex$_{(1)}$.} \leqno (7) $$ Again, the 2-point vertex$_{(0)}$
does not appear in the case of a massless theory.

\noindent There is moreover an overall normalization factor which
depends upon the conventions for the choice of the Haar measure in
$d$-dimensional space. We shall normalize the Haar measure so
that, $$ \int d^d \, p \, \exp \, (-p^2) = \pi^{d/2} \, . \leqno
(8) $$ This introduces an overall factor of $(2\pi)^{-dL}$ where
$L$ is the loop number of the graph, i.e. the number of internal
momenta.

\smallskip

\noindent The integral (2) makes sense using the rules of dimensional
regularization (cf.~\cite{C} Chap.~4) provided the complex number $d$ is in
a neighborhood of $D=6$ and $d \ne D$.

\smallskip

\noindent If we let $\s$ be the external momenta structure of the graph
$\G$ we would like to define the bare value $U \, (\G)$ simply by evaluating
$\s$ on the test function (2) but we have to take care of two
requirements. First we want $U \, (\G)$ to be a pure number, i.e. to be a
dimensionless quantity. In order to achieve this we simply multiply $\lgl
\s , U_{\G} \rgl$ by the appropriate power of $\mu$ to make it
dimensionless.

\smallskip

\noindent The second requirement is that, for a graph with $N$ external legs
we should divide by $g^{N-2}$ where $g$ is the coupling constant.

\smallskip

\noindent We shall thus let:
$$
U \, (\G) = g^{(2-N)} \, \mu^{-B} \, \lgl \s , U_{\G} \rgl \leqno (9)
$$
where $B = B \, (d)$ is the dimension of $\lgl \s , U_{\G} \rgl$.

\smallskip

\noindent Using (3)-(7) this dimension is easy to compute. One can remove
all 2-point vertices from the graph $\G$ without changing the dimension of
$U_{\G}$ since removing such a vertex removes an internal line and a factor
(6), (7) which by (3) does not alter the dimension. Thus let us assume that
all vertices of $\G$ are 3-point vertices. Each of them contributes by
$(3-d/2)$ to the dimension and each internal line by $-2$, and each loop by
$d$ (because of the integration on the corresponding momenta). This gives
$$
\dim \, (U_{\G}) = \left( 3 - \frac{d}{2} \right) \, V - 2 \, I + d \, L \,
. \leqno (10)
$$
One has $L = I - V + 1$, so the coefficient of $d$ in (10) is $I -
\frac{3}{2} \, V + 1 = 1 - \frac{N}{2}$. (The equality $N = 3 \, V - 2 \, I$
follows by considering the set of pairs $(x,y)$ where $x \in \G^{(0)}$ is a
vertex, $y \in \G^{(1)}$ is an internal line and $x \in y$.) The constant
term is $3 \, V - 2 \, I = N$ thus,
$$
\dim \, (U_{\G}) = \left( 1 - \frac{N}{2} \right) \, d + N \, . \leqno (11)
$$
We thus get,
$$
B = \left( 1 - \frac{N}{2} \right) \, d + N + \dim \, \s \, . \leqno (12)
$$
Thus (11) and (12) are valid for arbitrary 1PI graphs (connected) since both
sides are unchanged by the removal of all two point vertices.

\smallskip

\noindent To understand the factor $g^{2-N}$ in (9) let us show
that the integer valued function, $$ \hbox{Order} \, \G = V_3 -
(N-2) \leqno (13) $$ where $V_3$ is the number of 3-point vertices
of the 1PI connected graph $\G$, and $N$ the number of its
external lines, does define a grading in the sense of (31) section
2. We need to check that, $$ \hbox{Order} \, (\g) + \hbox{Order}
\, (\G / \g) = \hbox{Order} \, \G \leqno (14) $$ using the
notations of section 2, with $\g$ connected. If $\g$ is a self
energy graph one has Order $\g = v_3$ which is the number of
3-point vertices removed from $\G$ in passing to $\G / \g$, thus
(14) follows. If $\g$ is a vertex graph, then Order $\g = v_3 - 1$
which is again the number of 3-point vertices removed from $\G$ in
passing to $\G / \g$ since $\g$ is replaced by a 3-point vertex in
this operation. Thus (14) holds in all cases.

\smallskip

\noindent Thus we see that the reason for the convention for powers of $g$
in (9) is to ensure that $U \, (\G)$ is a monomial of degree Order $(\G)$ in
$g$.

\smallskip

\noindent We extend the definition (9) to disjoint unions of 1PI
graphs $\G_j$ by, $$ U \, (\G = \cup \, \G_j) = \Pi \, U \, (\G_j)
\, . \leqno (15) $$ One can of course write simple formulas
involving the number of external legs and the number of connected
components of $\G$ to compare $U \, (\G)$ with $\lgl \s , U_{\G}
\rgl$ as in (9).

\smallskip

\noindent Before we state the main result of this paper in the
form of Theorem~4 below let us first recall that if it were not
for the divergences occuring at $d=D$, one could give perturbative
formulas for all the important physical observables of the QFT in
terms of sums over Feynman graphs. It is of course a trivial
matter to rewrite the result in terms of the $U \, (\G)$ defined
above and we shall just give a few illustrative examples.

\smallskip

\noindent The simplest example is the effective potential which is an
ordinary function,
$$
\phi_c \ra V (\phi_c) \leqno (16)
$$
of one variable traditionally noted $\phi_c$.

\smallskip

\noindent The $n^{\rm th}$ derivative $V^{(n)} \, (0)$ is just
given as the sum, with $S \, (\G)$ the symmetry factor of $\G$, $$
V^{(n)} \, (0) = \sum_{\G} \, 1 / S \, (\G) \, \lgl \s_0 , U_{\G}
\rgl \leqno (17) $$ where $\s_0$ is evaluation at 0 external
momenta, and where $\G$ varies through all 1PI graphs with $n$
external momenta and only 3-point interaction
vertices.\footnote{The reader should not forget that we committed
ourselves to an Euclidean metric, so that appropriate Wick
rotations are necessary to compare with results obtained  in
Minkowski space.} We thus get $$ V^{(n)} \, (0) = g^{n-2} \,
\mu^{n-d \, \left( \frac{n}{2} - 1 \right)} \sum_{\G} \,
\frac{1}{S \, (\G)} \, U \, (\G) \leqno (18) $$ but this
expression is meaningless at $d=D$ which is the case of interest.

\smallskip

\noindent If instead of evaluating at 0 external momenta one keeps the
dependence on $p_1 , \ldots , p_n$ one obtains the expression for the
effective action,
$$
\Lb = \sum_n \, \frac{1}{n!} \, \int d^d \, x_1 \ldots d^d \, x_n \,
\Lb^{(n)} (x_1 , \ldots , x_n) \, \phi_c (x_1) \ldots \phi_c (x_n) \leqno
(19)
$$
where $\Lb^{(n)} \, (p_1 , \ldots , p_n)$ is given by (18) evaluated at
external momenta given by the $p_j$'s.

\smallskip

\noindent For expressions which involve connected diagrams which are not
1PI, such as the connected Green's functions one just needs to express the
bare value of a graph $U_{\G} \, (p_1 , \ldots , p_N)$ in terms of the bare
values of the 1PI components which drop out when removing internal lines
which carry a fixed value (depending only on $p_1 , \ldots , p_N)$ for their
momenta.

$$ \hbox{ \psfig{figure=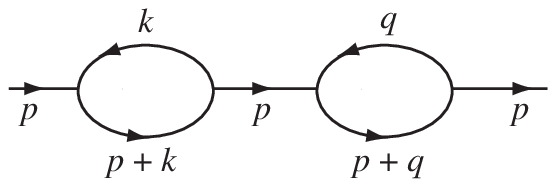} } $$

\noindent In this example we get $U_{\G} \, (p,-p) = U_{\G_1} (p,-p) \,
U_{\G_2} (p,-p)$ where $\G_j$ is the one loop self energy graph.

\smallskip

\noindent Similarly for expressions such as the Green's functions
which involve diagrams which are not connected one simply uses the
equality $$ U_{\G_1 \, \cup \, \G_2} = U_{\G_1} \, U_{\G_2} \leqno
(20) $$ where $\G_1 \cup \G_2$ is the disjoint union of $\G_1$ and
$\G_2$. In all cases the graphs involved are only those with
3-point interaction vertices and the obtained expressions only
contain finitely many terms of a given order in terms of the
grading (13). If it were not for the divergences at $d=D$ they
would be the physical meaningful candidates for an asymptotic
expansion in terms of $g$ for the value of the observable.

\smallskip

\noindent Let us now state our main result:

\medskip

\noindent {\tencsc Theorem 4.} a) {\it There exists a unique loop $\g (z)
\in G$, $z \in \Cb$, $\vert z - D \vert < 1$, $z \ne D$ whose
$\G$-coordinates are given by $U \, (\G)_{d=z}$.}

\smallskip

\noindent b) {\it The renormalized value of a physical observable
$\Oc$ is obtained by replacing $\g \, (D)$ in the perturbative
expansion of $\Oc$ by $\g_+ \, (D)$ where $$ \g \, (z) = \g_- \,
(z)^{-1} \ \g_+ \, (z) $$ is the Birkhoff decomposition of the
loop $\g \, (z)$ on any circle with center $D$ and radius $< 1$.}

\medskip

\noindent {\it Proof.} To specify the renormalization we use the
graph by graph method \cite{BP} using dimensional regularization
and the minimal substraction scheme. We just need to concentrate
on the renormalization of 1PI graphs $\G$. We shall use the
notations of \cite{C} to make the proof more readable. Our first
task will be to express the Bogoliubov, Parasiuk and Hepp
recursive construction of the counterterms $\G \ra C \, (\G)$ and
of the renormalized values of the graphs $\G \ra R \, (\G)$, in
terms of the Hopf algebra $\Hc$.

\smallskip

\noindent We fix a circle $C$ in $\Cb$ with center at $D=6$ and radius $r <
1$ and let $T$ be the projection
$$
T : \Ac \ra \Ac_- \leqno (21)
$$
where $\Ac$ is the algebra of smooth functions on $C$ which are meromorphic
inside $C$ with poles only at $D=6$, while $\Ac_-$ is the subalgebra of
$\Ac$ given by polynomials in $\frac{1}{z-6}$ with no constant term. The
projection $T$ is uniquely specified by its kernel,
$$
{\rm Ker} \, T = \Ac_+ \leqno (22)
$$
which is the algebra of smooth functions on $C$ which are holomorphic inside
$C$.

\smallskip

\noindent Thus $T$ is the operation which projects on the pole part of a
Laurent series according to the MS scheme. It is quite important
(cf.~\cite{C} p.~103 and 6.3.1 p.~147) that $T$ is only applied to
dimensionless quantities which will be ensured by our conventions in the
definition (9) of $U \, (\G)$.

\smallskip

\noindent Being the projection on a subalgebra, $\Ac_-$, parallel to a
subalgebra, $\Ac_+$, the operation $T$ satisfies an equation of
compatibility with the algebra structure of $\Ac$, the multiplicativity
constraint (\cite{K3}),
$$
T \, (x \, y) + T(x) \, T(y) = T \, (T(x) \, y + x \, T(y)) \qquad \fl \, x
, y \in \Ac \, . \leqno (23)
$$
(By bilinearity it is enough to check it for $x \in \Ac_{\pm}$, $y \in
\Ac_{\pm}$.)

\smallskip

\noindent We let $U$ be the homomorphism,
$$
U : \Hc \ra \Ac \leqno (24)
$$
given by the unrenormalized values of the graphs as defined in (9) and (15).
It is by construction a homomorphism from $\Hc$ to $\Ac$, both viewed as
algebras.

\smallskip

\noindent Let us start the inductive construction of $C$ and $R$.
For 1PI graphs $\G$ without subdivergences one defines $C \, (\G)$
simply by, $$ C \, (\G) = -T \, (U \, (\G)) \, . \leqno (25) $$
The renormalized value of such a graph is then, $$ R \, (\G) = U
\, (\G) + C \, (\G) \, . \leqno (26) $$ For 1PI graphs $\G$ with
subdivergences one has, $$ C \, (\G) = -T \, (\ov R \, (\G))
\leqno (27) $$ $$ R \, (\G) = \ov R \, (\G) - T \, (\ov R \, (\G))
\leqno (28) $$ where the $\ov R$ operation of Bogoliubov, Parasiuk
and Hepp prepares the graph $\G$ by taking into account the
counterterms $C \, (\g)$ attached to its subdivergences. It is at
this point that we make contact with the coproduct (II.6) of the
Hopf algebra $\Hc$ and claim that the following holds, $$ \ov R \,
(\G) = U \, (\G) + \sum_{\g \sbsneq \G} C (\g) \ U (\G / \g) \, .
\leqno (29) $$ The notations are the same as in II.6. This formula
is identical to 5.3.8~b) in \cite{C} p.~104 provided we carefully
translate our notations from one case to the other. The first
point is that the recursive definition (27) holds for 1PI graphs
and $\g = \cup \, \g_j$ is a union of such graphs so in (29) we
let, $$ C \, (\g) = \Pi \ C (\g_j) \, . \leqno (30) $$ This agrees
with (5.5.3) p.~110 in \cite{C}.

\smallskip

\noindent The second point is that our $C \, (\g)$ is an element of $\Ac$,
i.e. a Laurent series, while the counterterms used in \cite{C} are in
general functions of the momenta which flow through the subdivergence.
However since the theory is renormalizable we know that this dependence
corresponds exactly to one of the three terms in the original Lagrangian.
This means that with the notations of II.6 we have,
$$
C_{\g} (\G) = \sum_{(i)} C (\g_{(i)}) \ U (\G / \g_{(i)}) \leqno (31)
$$
where $C_{\g} (\G)$ is the graph with counterterms associated to the
subdivergence $\g$ as in Collins 5.3.8~b).

\smallskip

\noindent To check (31) we have to check that our convention (9) is correct.
As we already stressed the power of the unit of mass $\mu$ is chosen
uniquely so that we only deal for $U$, $C$ and $R$ with dimensionless
quantities so this is in agreement with \cite{C} (2)p.~136. For the power of the coupling constant $g$ it
follows from (14) that it defines a grading of the Hopf algebra so that all
terms in (31) have the same overall homogeneity. There is still the question
of the symmetry factors since it would seem at first sight that there is a
discrepancy between the convention (4) p.~24 of \cite{C} and our convention
in (9).
However a close look at the conventions of \cite{C} (cf.~p.~114) for the
symmetry factors of $C_{\g} (\G)$ shows that (31) holds with our
conventions. We
have thus checked that (29) holds and we can now write the BPH recursive
definition of $C$ and $R$ as follows, replacing $\ov R$ by its value (29)
for a
1PI graph $\G$ in (27), (28),
$$
C \, (\G) = -T \left( U (\G) + \sum_{\g \sbsneq \G} C (\g) \, U (\G / \g)
\right) \leqno (32)
$$
$$
R \, (\G) = U (\G) + C (\G) + \sum_{\g \sbsneq \G} C (\g) \, U (\G / \g) \,.
\leqno (33)
$$
We now rewrite both formulas (32), (33) in terms of the Hopf algebra $\Hc$
without using the generators $\G$. Let us first consider (32) which together
with (30) uniquely determines the homomorphism $C : \Hc \ra \Ac$. We claim
that
for any $X \in \Hc$ belonging to the augmentation ideal
$$
\wt \Hc = {\rm Ker} \, \ov e \leqno (34)
$$
one has the equality,
$$
C(X) = -T (U(X) + \sum \, C (X') \, U (X'')) \leqno (35)
$$
where we use the following slight variant of the Sweedler notation for the
coproduct $\D \, X$,
$$
\D \, X = X \ot 1 + 1 \ot X + \sum \, X' \ot X'' \ , \quad X \in \wt \Hc
\leqno (36)
$$
and where the components $X'$, $X''$ are of degree strictly less than the
degree
of $X$.
\smallskip
\noindent To show that (35) holds for any $X \in \wt \Hc$ using (30) and
(32)
one defines a map $C' : \Hc \ra \Ac$ using (35) and one just needs to show
that
$C'$ is multiplicative. This is done in \cite{CK} \cite{K3} but we repeat the
argument here for the sake of completeness. One has, for $X , Y \in \wt \Hc$
$$
\ldisplaylinesno{
\D (XY) = XY \ot 1 + 1 \ot XY + X \ot Y + Y \ot X + XY' \ot Y'' \, + &(37)
\cr
Y' \ot X Y'' + X' Y \ot X'' + X' \ot X'' Y + X' Y' \ot X'' Y'' \, . \cr
}
$$
Thus using (35) we get
$$
\ldisplaylinesno{
C' (XY) = -T (U (XY)) - T (C' (X) \, U(Y) + C' (Y) \, U(X) \, + &(38) \cr
C' (XY') \, U (Y'') + C' (Y') \, U (XY'') + C' (X' Y) \, U(X'') \cr
+ \, C' (X') \, U (X'' Y) + C' (X' Y') \, U (X'' Y'')) \, . \cr
}
$$
Now $U$ is a homomorphism and we can assume that we have shown $C'$ to be
multiplicative, $C' (AB) = C' (A) \, C' (B)$ for $\deg A + \deg B < \deg X +
\deg Y$. This allows to rewrite (38) as,
$$
\ldisplaylinesno{
C' (XY) = -T (U (X) \, U(Y) + C' (X) \, U (Y) + C' (Y) \, U(X) &(39) \cr
+ \, C' (X) \, C' (Y') \, U (Y'') + C' (Y') \, U (X) \, U (Y'') + C' (X') \,
C'
(Y) \, U (X'') \cr
+ \, C' (X') \, U (X'') \, U (Y) + C' (X') \, C' (Y') \, U (X'') \, U (Y'')
\, .
\cr
}
$$
Let us now compute $C' (X) \, C' (Y)$ using the multiplicativity constraint
(23)
fulfilled by $T$ in the form,
$$
T(x) \, T(y) = -T (xy) + T (T(x) \, y) + T (x \, T(y)) \, . \leqno (40)
$$
We thus get,
$$
\ldisplaylinesno{
C' (X) \, C' (Y) = -T ((U(X) + C' (X') \, U (X'')) \, (U (Y) \, + &(41) \cr
C' (Y') \, U (Y'')) + T (T (U(X) + C' (X') \, U (X'')) \, (U(Y) \, + \cr
C' (Y') \, U (Y'')) + T ((U(X) + C' (X') \, U(X'')) \, T (U(Y) + C' (Y') \,
U
(Y''))) \cr
}
$$
by applying (40) to $x = U(X) + C(X') \, U(X'')$, $y = U(Y) + C (Y') \, U
(Y'')$. Since $T(x) = -C' (X)$, $T(y) = -C' (Y)$ we can rewrite (41) as,
$$
\ldisplaylinesno{
C' (X) \, C' (Y) = -T (U(X) \, U (Y) + C' (X') \, U (X'') \, U (Y) &(42) \cr
+ \, U(X) \, C' (Y') \, U(Y'') + C' (X') \, U (X'') \, C' (Y') \, U (Y''))
\cr
-T (C'(X) (U(Y) + C' (Y') \, U(Y'')) - T ((U(X) + C' (X') \, U (X'')) \, C'
(Y))
\, . \cr
}
$$
We now compare (39) with (42), both of them contain 8 terms of the form
$-T(a)$
and one checks that they correspond pairwise which yields the
multiplicativity
of $C'$ and hence the validity of (35) for $C = C'$.
\smallskip
\noindent We now have a characterization of $C$ independently of any choice
of
generators of $\Hc$ and we can rewrite (33) in intrinsic form too,
$$
R(X) = \lgl C \ot U , \D (X) \rgl \qquad \fl \, X \in \Hc \leqno (43)
$$
which can be checked using (33) and the multiplicativity of both sides of
(43).
\smallskip
\noindent It is convenient to use the notation $C \star \, U$ for the
homomorphism $\Hc \ra \Ac$ given by, $$ (C \star \, U) (X) = \lgl C
\ot U , \D (X) \rgl \qquad \fl \, X \in \Hc \, . \leqno (44) $$ We
are now ready to check that (43) gives the Birkhoff decomposition
of the loop $\g \, (z)$, $z \in C$ of elements of the group $G$ of
section 2, associated to the homomorphism, $$ U : \Hc \ra \Ac \, .
\leqno (45) $$ The precise definition of $\g$ is as follows. Each
complex number $z \in C$ defines a character of the algebra $\Ac$
given by, $$ \chi_z (f) = f(z) \qquad \fl f \in \Ac \leqno (46) $$
which makes sense since $f$ is smooth on the curve $C$. Thus
$\chi_z \circ U$ is a character of $\Hc$ and hence (cf.~section 2)
an element of $G$, $$ \g (z) = \chi_z \circ U \qquad \fl \, z \in
C \, . \leqno (47) $$ Next we can similarly define two other loops
with values in $G$, namely, $$ \g_- (z) = \chi_z \circ C \ , \
\g_+ (z) = \chi_z \circ R \qquad \fl \, z \in C \, . \leqno (48)
$$ The multiplicativity of both $C$ and $R$, $\Hc \ra \Ac$ ensures
that we are dealing with $G$-valued loops.
\smallskip
\noindent The equality (43) just means,
$$
\g_+ (z) = \g_- (z) \, \g (z) \qquad \fl \, z \in C \leqno (49)
$$
since (44) is the same as the operation of pointwise product in $G$ for
$G$-valued loops.
\smallskip
\noindent It remains to check that $\g_{\pm}$ extends to a $G$-valued map
holomorphic in $C_+$.
\smallskip
\noindent By (35) one has $C(\wt \Hc ) \sbs \Ac_-$ and every $z
\in C_-$ defines using (46) a character on $\Ac_-$ with
$\chi_{\ify} = 0$ the trivial character. It thus follows that
$\g_-$ extends to a $G$-valued map holomorphic on $C_-$ and such
that, $$ \g_- (\ify) = 1 \, . \leqno (50) $$ Similarly, by (35)
one has $T((C \star\, U) (X)) = 0$ so that $R (\wt \Hc ) \sbs
\Ac_+ = {\rm Ker} \, T$. Since every $z \in C_+$ defines using
(46) a character of $\Ac_+$ we see that $\g_+$ extends to a
$G$-valued map holomorphic in $C_+$. This shows that (49) gives
the Birkhoff decomposition of $\g$ and that the renormalized value
$R(\G)$ of any 1PI graph is simply obtained by replacing the ill
defined evaluation $\g (D)$ by $\g_+ (D)$.~\xx

\medskip

\noindent Again, a few remarks are in order. First, the above
decomposition singles out the use of minimal subtraction together
with Dim.Reg.~as a favoured approach. From here, one can find the
relation to other schemes using the methods of \cite{K3}. In
\cite{CKnew} we will discuss the global nature of the group $G_c$,
its relation with diffeomorphism groups and with the
renormalization group.
We shall also discuss the relation between the
quantized calculus and the
reduction to first order
 poles implicitly allowed by the above combinatorial structure.
\medskip

\noindent Finally, the reader might expect that the relation to the
Riemann-Hilbert problem indicates the presence of a differential
equation in $z$ which relates the $z$ dependence of counterterms
and renormalized Green's functions to derivations on the Hopf
algebra. Such a differential equation can be given and the relation of monodromy
 to anomalous dimensions will be discussed in \cite{CKnew}
as well.
\section*{Acknowledgements}
D.K.~thanks the Clay Mathematics Institute for support during a
stay at Lyman Laboratories, Harvard University, and is grateful to
the DFG for a Heisenberg Fellowship.

\end{document}